\documentstyle[11pt,newpasp,twoside,psfig]{article}
\index{summary}
\index{instructions}
\index{template}

\def\edcomment#1{\iffalse\marginpar{\raggedright\sl#1\/}\else\relax\fi}
\reversemarginpar

\begin{document}
\title{Low-Mass Neutron Stars as Anomalous Pulsars}
 \author{Frederick D. Seward}
\affil{Smithsonian Astrophysical Observatory, 60 Garden Street,
 Cambridge, MA 02138}

\begin{abstract}
A neutron star with mass close to the lower limit might be a
reasonable model for some anomalous pulsars.  Emission is
thermal. X-ray luminosity is high. Spatial velocity can be high.
Since the radius is predicted to be large, the magnetic field 
calculated 
for spin-down is lower than that required by the magnetar model.

\end{abstract}

\vspace{-.3in}
\section{Introduction}

Almost every pulsar has precisely measured
period, $P$, and the rate of change of period with time, $\dot{P}$.  The
gross properties of these objects are well illustrated by the
$P-\dot{P}$
 diagram which now shows rotational data for $\sim 1200$ isolated
(as far as we know) pulsars.  The accepted model, a rotating,
strongly magnetized, neutron star, gives a plausible explanation for
the behavior and evolution of most of these objects.

Approximately 10 objects,
having both large $P$ and $\dot{P}$, form a loose group in the upper right 
corner.  Here the dipole model predicts large values for 
the magnetic moment, $M$, and the surface field, $B$.
Indeed, the energy in the predicted magnetic field exceeds the
 rotational energy in this part of the diagram.  These objects are the 
Anomalous X-ray Pulsars (AXP) and the the Soft Gamma-ray Repeaters (SGR)
which we refer to collectively as anomalous pulsars.  Also for these
pulsars the observed X-ray
luminosity also greatly exceeds the rate of loss of rotational energy.

Some anomalous pulsars are associated with supernova remnants, and the short
lifetimes indicated by $P/2\dot{P}$ seem appropriate.  It
has been proposed (Thompson \& Duncan 1996) that energy is supplied by
decay of the
very-strong magnetic field.  The interaction of the strong field with
the crust of the neutron star is also invoked to explain the $\gamma$-ray
bursts observed from SGR.  AXP/SGRs are also called ``magnetars'' in the
literature.  It is assumed that the slowing torque is
largely electromagnetic. However,
the grouping of anomalous pulsars in the upper right of the $P-\dot{P}$ diagram,
suggests that something different is happening rather than an
extension of the dipole model.  We here explore the possibility that the 
anomalous pulsars are low-mass
neutron stars -- objects with mass considerably smaller than the
generally-accepted mass of 1.4 M$_\odot$. 

\section{Neutron Star Structure}

In theory, neutron stars can exist having masses between $\approx$ 0.2
M$_{\odot}$ and $\approx$ 3.0 M$_{\odot}$.  The addition of mass
at the upper limit causes formation of a black hole.  The removal of
mass at the lower limit causes nuclei in the crust to fission
and beta decay with release of energy and consequent disruption of the
star (Colpi, Shapiro, Teukolsky, 1991).  

The only mass measurements are for neutron stars in binary systems.  
The masses of 19 radio pulsars, deduced from measured orbital
parameters and characteristics of the companion star
are all compatible with a mass of $ 1.35\pm 0.04$ M$_{\odot}$
(Thorsett \& Chakrabarty, 1999), but
 uncertainty of the measurements for 2 systems allow
masses slightly below 1.0 M$_{\odot}$.  The masses of 7
X-ray emitting accretion-powered pulsars in binary systems are
all
compatible with a mass of 1.4 M$_{\odot}$, but larger uncertainties 
allow masses ranging from 0.5 to 2.5 M$_{\odot}$ (van Kerkwijk et al 1995).

\begin{figure}[t]
\centerline{\psfig{file=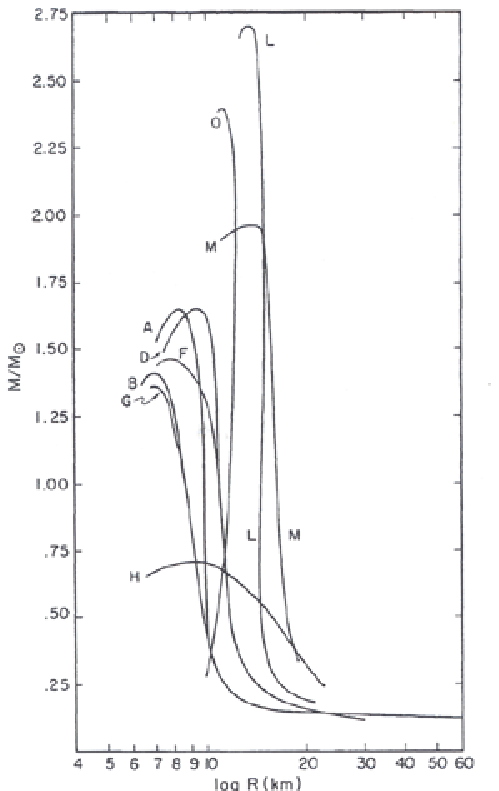,height=3in}\psfig{file=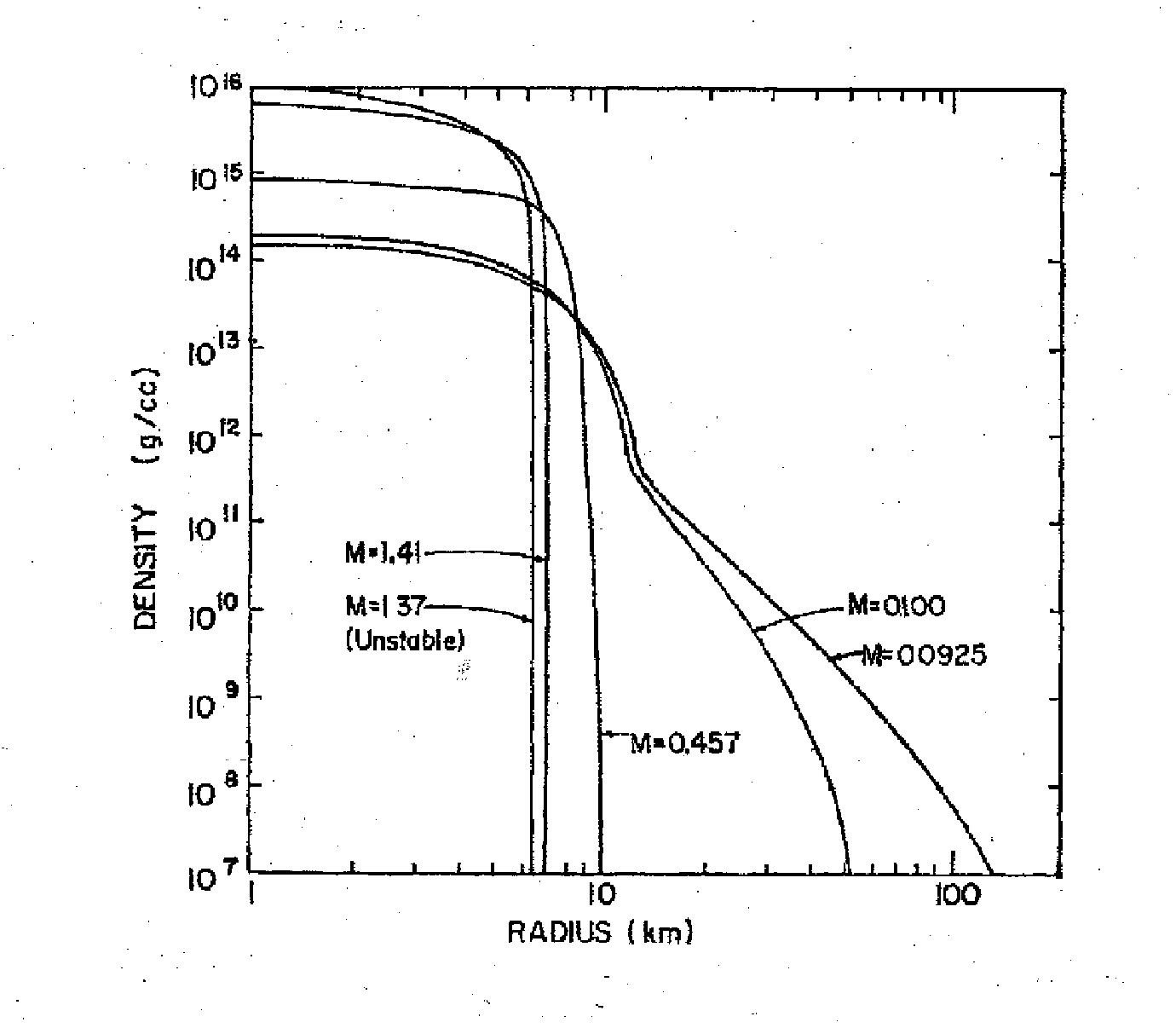,height=3.2in}}
\caption{\it (a)Neutron star mass-radius relation for several eq. of
  state.  (b)Density vs radius for neutron stars of different masses.}
\end{figure}

Since gravitational collapse of a massive star 
in a type II supernova is thought to produce an $\approx 1.4$
M$_{\odot}$ compact object, the general belief is that all neutron
stars are so formed and have this mass.  Larger masses might be achieved
through accretion by neutron stars in binary systems
 but there is presently no
evidence for masses $\leq$ 1.3 M$_{\odot}$; so the subject of this paper
is speculative.

Neutron star properties, including the lower-mass limit, are
dependent on the equation of state which is unknown for the
high density at the center of a neutron star.
Arnett \& Bowers (1977) constructed neutron star models for 15
equations of state (Figure 1a).  More recent models are given by Lattimer \&
Prakash (2000).  Lower limits of 
0.2 - 0.4 M$_{\odot}$ are calculated
for most models.  ``Soft'' equations of state predict an almost
constant radius of between 8 and 15 km (dependent on eq. of
state) for masses 0.5-1.4 M$_{\odot}$.  Then, as the 
mass approaches the lower limit, the radius, $R$, increases rapidly.  
A ``stiff'' equation of state predicts a more linear
dependence of radius on mass.  


Figure 1b, from Baym et al (1971), shows density profiles calculated using one
eq. of state and 5 different masses.  The 1.4 M$_{\odot}$ star
has a radius of 7 km, the 0.5 M$_{\odot}$ star a radius of 10 km, and
the 0.1 M$_{\odot}$ star a radius of 50 km.  As a point of interest, 
the 0.1 M$_{\odot}$ star is completely solid; there is no liquid core.

\section{Cooling}

A neutron star formed in a supernova explosion should be born hot and 
will cool
with time.  Cooling is dominated by neutrino emission from the core for
$\approx$ $10^4$ years, then by photon emission from the surface.
Initially the core cools rapidly and the surface remains hot until 
the heat is conducted to the core.  

Figure 2, taken from Page and Applegate (1992), shows cooling
curves calculated for neutron stars of different masses and includes a
direct URCA process which cools the core rapidly.  Below a certain
mass the central density is low enough so that the direct URCA process
no longer occurs.  The time for the cooling wave to reach the
surface increases as the mass decreases because the crust gets
thicker.  

\makebox[3in][l]
{\psfig{file=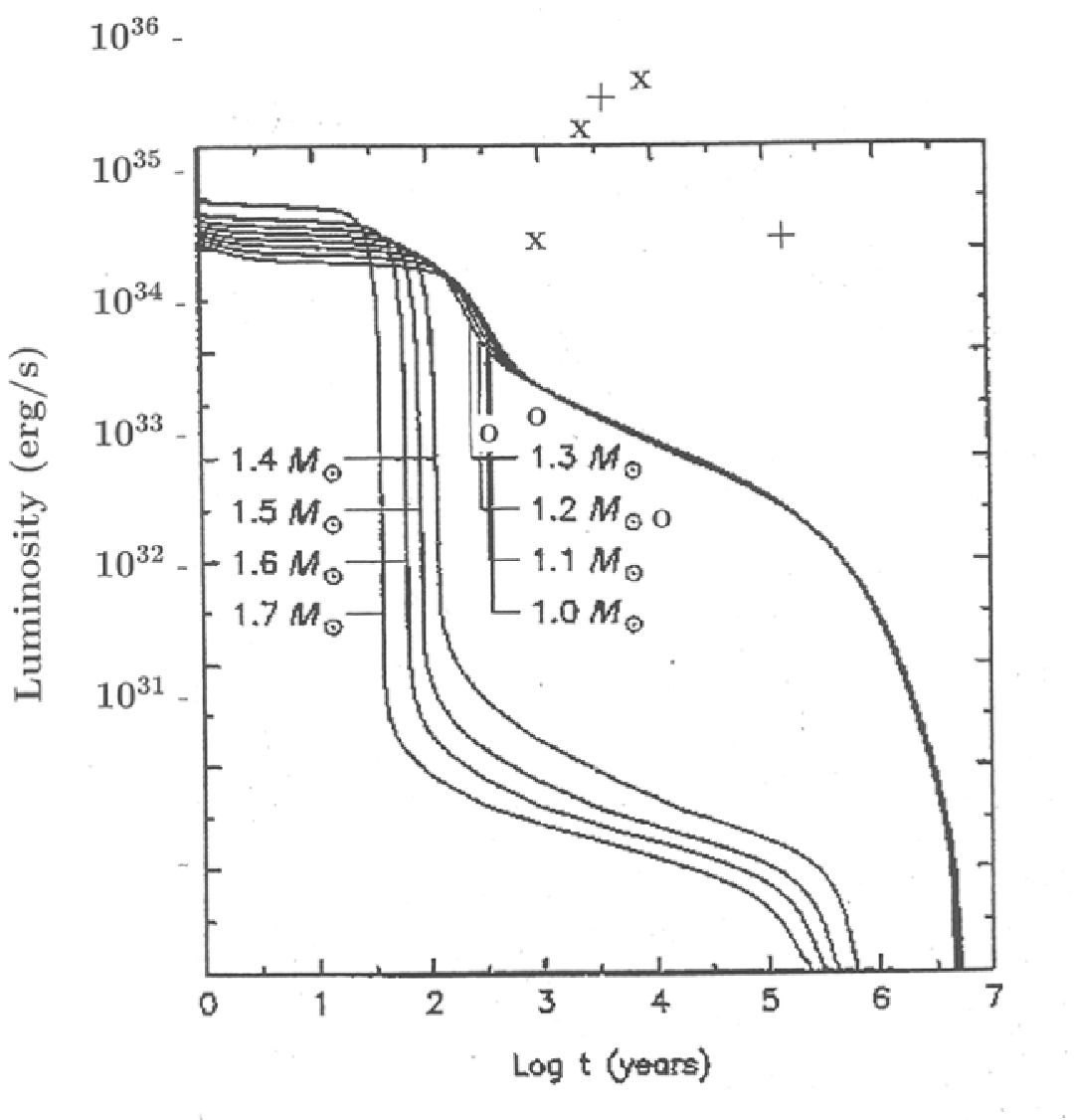,width=3in,angle=-0.8}}
\parbox[b]{2in}{Figure 2. {\it Luminosity vs time for neutron stars
of different masses.
Circles below the highest curve show luminosities of 3 young
pulsars: the central sources
in the remnants: Cas A, 3C 58, and the Vela remnant.
Data points above the curve are for 2 AXP and 3 SGR. 
Ages are uncertain by at least
a factor of 2 and distances are not well known 
except for one SGR in the Large Magellanic cloud.}}

\vspace{0.2in}
If a neutron star has low central 
density (and consequent low mass), early neutrino cooling will 
be greatly reduced.
There will be no ``exotic'' cooling processes and the crust 
will be thick.  Even if there is some cooling of
the core, the surface will remain hot until cooled by photon
emission. The larger
radius expected from a low-mass star will also help to achieve
a high luminosity. 
At an age of a few thousand years, the star must have
thermal luminosity in excess of $10^{35}$ erg s$^{-1}$.

As a matter of interest, if thermal conductivity is greater 
along the field lines, and if the
core cools first, phase-resolved spectra might show cool spots 
at the magnetic poles - quite different than
the hot poles expected from accretion.

\section{Magnetic Field}

The spindown rate, $\dot{P}$, of anomalous pulsars is erratic,
demonstrating that another process, probably a wind, is
responsible for much of the torque (Marsden et al, 1999).  Certainly
some magnetic field is required to account for the pulsations but
characteristic ages calculated using the dipole model (and used to
place points in Figure 2) are not expected to be reliable.

One difficulty with the magnetar model is the high value of magnetic
field at the surface (eg. $10^{15}$ Gauss for SGR 1900+14).  
Equating  the loss
of rotational energy to the rate of dipole radiation gives a value for
the dipole moment, $M$.  Since the external field of a dipole along the axis
is $B_{p}=M/2R^3$, the surface field at the pole can be expressed as:

\begin{displaymath}
                B_p = [3c^{3}IP\dot{P}/8\pi^{2}]^{1/2}[R]^{-3}
\end{displaymath}

Given $M$, a low-mass neutron star with large $R$ will have considerably
lower surface field than the usual ($R = 10$ km) model.  Also
the moment of inertia, $I$, roughly scales with mass (Arnett \& Bowers
1977).  If the
mass of the model is lowered until $R = 20$ km, the value of $B$ at the surface
will decrease  a factor of 10.  If the radius is 40 km, $B$ will be 100
times less than the usual model.  The total energy in the field will
also drop factors of 10 and 100 in these 2 examples.

\section{Other Considerations}

Some anomalous pulsars are thought to be high-velocity objects.  Less
energy is required to accelerate a low-mass star than a canonical
one.  This might be accomplished through asymmetric radiation
(Harrison \& Tademaru 1975) or during formation.  

The process by which low-mass neutron stars may be formed 
is unknown.  Gravitational collapse requires a core of $\sim 1
M_{\odot}$.  The collapse of a rotating system with magnetic field might,
in spite of gravitational radiation, produce smaller fragments.
It will be difficult, however, to identify a low-mass isolated 
neutron star with
present observational techique.  The surface
and emission process are not well enough 
understood to derive the radius from spectral data.
  
This work was supported by NASA contract NAS8-39073.

\end{document}